\documentclass[journal,12pt,onecolumn,draftclsnofoot,]{IEEEtran}

\usepackage{amsmath}
\usepackage{graphicx}
\usepackage{hhline}
\usepackage[space]{cite}
\usepackage{notoccite}
\usepackage[thinlines]{easytable}
\usepackage{multirow}

\begin{document}

\title{Mixed-Effect Modeling for Longitudinal Prediction of Cancer Tumor}
\author{Fatemeh~Nasiri and Oscar~Acosta-Tamayo}
\maketitle

\begin{abstract}
In this paper, a mixed-effect modeling scheme is proposed to construct a predictor for different features of cancer tumor. For this purpose, a set of features is extracted from two groups of patients with the same type of cancer but with two medical outcome: 1) survived and 2) passed away. The goal is to build different models for the two groups, where in each group, patient-specified behavior of individuals can be characterized. These models are then used as predictors to forecast future state of patients with a given history or initial state. To this end, a leave-on-out cross validation method is used to measure the prediction accuracy of each patient-specified model. Experiments show that compared to fixed-effect modeling (regression), mixed-effect modeling has a superior performance on some of the extracted features and similar or worse performance on the others.

\end{abstract}

% ========================================================
% ========================================================
% ========================================================

\label{sec:framework}
\section{Introduction}
\label{introduction}

With the increasing prevalence of the signal processing applications in the medical science, the need for sophisticated automated processing tools is more perceived. One group of these applications deals with different imaging systems (e.g. MRI, CT, X-ray etc.), where one can derive useful information by studying big and diverse datasets of images taken from different patients in different stages of a disease. However, this diversity and plurality of data is helpful only if a proper and accurate automated tool is designed to process them.

Longitudinal study and disease tracking using medical imaging is one of the most important and effective processes during the medical treatment. Generally speaking, this procedure usually requires images regularly taken from patients' organ(s) involving in disease evolution. The information in these images hopefully guides to make better estimations regarding the future state of the organs and observe the impact of different drugs in different situations. For instance, one might be interested to discover the impact of a specific dosage of an anti-cancer drug on the size of the tumor through the time. Therefore, the question is how to benefit from the latent information in the temporally diverse set of medical images taken from different patients.

Statistical modeling and prediction is a practical and effective solution to address the above problem. The main idea is to exploit the repetitive behavior of the organ/tumors in response to a new condition (e.g. drug injection, aging etc.) and derive an accurate pattern. Like other similar machine learning and statistical modeling problems, this task requires a diverse image dataset with a reasonable size taken in different conditions. However, it has always been costly and more importantly dangerous to take excessive images from individual patients to specifically model their disease. The most feasible solution to this problem is to combine image datasets from different patients with the same disease condition and produce a rather general model. However, this simplification would compromise accuracy of the model due to the patient-specific behaviors of the treatments. 

In this paper, first a set of longitudinal related features from lung tumor are introduced and extracted. Then, a non-linear mixed-effect modelization scheme is proposed to construct predictors for each feature of each patient. The performance of these predictors are evaluated by a cross-validation algorithm.

The rest of this paper is organized as follows. In section \ref{seq:framework}, we first describe the key elements of a mixed-effect model with some examples in the longitudinal studies. Section \ref{sec:modelization} explains the main steps of the proposed mixed-effect modeling. In section \ref{sec:results}, we discuss the performance of the proposed mixed-effect modeling compared to fixed-effect modeling in terms of prediction accuracy and finally, section \ref{sec:conclusion} concludes the paper.
% =======================================
% =======================================
% =======================================
% =======================================
% =======================================
\section{General framework of a prediction system using mixed effect model}
\label{seq:framework}
In this section, we first explain principles of the mixed-effect modeling and its necessity in the modeling of a complex medical treatment process. Then, the main elements of a prediction system based on the mixed-effect modeling and their different choices are discussed.

\subsection{Principles of the mixed-effect modeling}
Basically, mixed-effect modeling or mixed modeling is a flexible statistical technique to exploit regularity in the pattern of a phenomenon which consists of both \textit{fixed} and \textit{random} effects \cite{nasiri2017review}. Herein, the term \textit{effect} refers to a system parameter that somehow impact the prediction value. This impact can be either non-random (i.e. fixed effects) or random (i.e. random effect) \cite{lindstrom1988newton} [cite mixed effect.pdf]. In matrix notation, a mixed-effect model can be represented as eq. \ref{eq:mixedmodelnotation}:

\begin{equation}
\label{eq:mixedmodelnotation}
y(f)=f(\beta)+g(\gamma)+\varepsilon,
\end{equation}
where $f(\beta)$ and $g(\gamma)$ are fixed and random effects of the model and $\varepsilon$ is the prediction error.

\subsection{Key elements of a mixed effect modeling system}
By taking a look at the longitudinal studies with medical image processing, one can spot a few common essential elements. Here we briefly describe these elements \cite{ribba2014review}.
\subsubsection{Tracked features}
Depending on the disease under study, the tracked feature can be different. In the problems dealing with different types of cancers, the objective is to observe and model the growth in the size of tumor. ``Sum of longest diameters'' and ``Mean tumor diameter'' are two size-related features that can fairly represent the status of the tumor and also can be helpful in tracking it. According to the Response Evaluation Criteria in Solid Tumors (RECIST) \cite{tsuchida2001response}, these features are allowed to be measured and tracked for on a limited set of organs associated with lesions.

However in other problems than the tomography, one might aim at monitoring more complicated features of the organ e.g. shape, displacement etc. For this purpose, some of the researches directly deal with pixels/voxels of the medical images taken from patients. In the simplest scenario, one might use each pixel/voxel of the image as one independent feature to track. More advanced approaches apply so-called feature reduction techniques such as Principle Component Analysis (PCA) to shrink the feature space \cite{rios2017population}.

In some applications, pixel/voxel level feature extraction requires pre-processing steps in order to improve accuracy of the extracted information. This step may include various image quality enhancement and restoration algorithms \cite{abdoli2015gaussian, katsaggelos2012digital, lee1980digital}.

There are also some other features that are used for tracking an organ's status through the time. For instance, in a research for studying the tumor growth rate, the Prostate-Specific Antigen (PSA) was tracked in order to monitor the prostate cancer status \cite{stein2010tumor}.

\subsubsection{Mixed-effect modelisation}
In the design of a mixed-effect model, there are different choices to make. One important choice is the mathematical expression used in modelisation. There are mainly two types of mixed-effect models in the literature: 
\begin{itemize}
\item Algebraic equations with the general form of:
\begin{equation}
y(t)=y_0+e^{-d.t}+g.t.
\end{equation} 
\item Differential equation with the general form of:
\begin{equation}
\frac{dy}{dt}=y_0+e^{-d.t}+g.t.
\end{equation}
\end{itemize}
In both equations, $y$ is the tracked feature, $t$ is time and $d$ is the model parameter. In this equation, the term exponential term of $e^{-d.t}$ is the random effect and the the term $g.t$ is the fixed effect in the model.

Once the mathematical expression is decided, one should decide about the number and the nature of random effects of the model. For example, here is a list of some popular random effects used in the literature of anti-cancer drug treatment:
\begin{itemize}
\item Drug-indudec decay of tumor
\item Net growth of tumor size
\item Tumor size nadir (the transition between decay and growth)
\item Drug concentration
\end{itemize}

\subsubsection{Response prediction}
As soon as the mixed effect model is trained with the training data, it can be used as a tool for prediction. This prediction generally includes estimation of future state of the tracked feature. The Expectation Maximization (EM) is one of the popular methods for addressing the response prediction problem.

% =======================================
% =======================================
% =======================================
% =======================================
% =======================================
% =======================================
\section{Modelization scheme}
\label{sec:modelization}
In this section, different steps of constructing patient-specified models are described in detail. The following sub-sections are presented in the order of modelization procedure, starting from raw medical images taken from patients and ending with patient-specified mixed-effect models that are capable of prediction.

\subsection{Feature extraction}
Similar to all statistical modeling and learning tasks, an essential element of the process is the feature selection. As explained, all features are extracted from radiology images.
\subsubsection{Population and imaging description}
The cases under this study are chosen from the patients of the Rennes Hospital diagnosed with lung cancer. From this population, two groups of deceased and survived  patients, each with 19 persons are randomly selected. 

All of the patients in the study have been through the same treatment but with different initial states. The treatment originally includes 6 weeks of regular imaging in order to track the situation of tumor. Each patient have been taken at least three images in this period. In other words, there are missing entries in the input data that are supposed to be properly treated during the modelization phase.

The first extracted feature is the volume of tumor. This criterion is measured by the number of voxels spatially belonging to the tumor region multiplied by the size of each volume which is defined by the imaging device. To have a precise modelization, spatial territory of each tumor is accurately determined by a radiologist. 

The second group of features relate to the tumor deformation through the time. For statistical presentation of tumor deformation, first we have performed a registration algorithm to fit the initial tumor shape on proper position in the next images. Then, the displacement of each voxel is calculated and formed a 1D vector. The Jacobian matrix of this vector is calculated to provide the second set of features. These features include mean, variance, shrewdness and kurtosis of the Jacobian matrix. The following equation shows the calculation of the Jacobian matrix from a vector $X$:

\begin{equation}
 {\displaystyle \mathbf {J} ={\begin{bmatrix}{\dfrac {\partial \mathbf {f} }{\partial x_{1}}}&\cdots &{\dfrac {\partial \mathbf {f} }{\partial x_{n}}}\end{bmatrix}}={\begin{bmatrix}{\dfrac {\partial f_{1}}{\partial x_{1}}}&\cdots &{\dfrac {\partial f_{1}}{\partial x_{n}}}\\\vdots &\ddots &\vdots \\{\dfrac {\partial f_{m}}{\partial x_{1}}}&\cdots &{\dfrac {\partial f_{m}}{\partial x_{n}}}\end{bmatrix}}}
\end{equation}

\subsection{Model design}
Once the features data are extracted, the next step is to derive a mathematical model that properly expresses the data properties. There are a handful of design choices in each data modelization task that need to be carefully made with respect to the nature of data. The major challenge in the longitudinal modelization is, as mentioned, the patient-specific behaviors. 

\begin{figure}
\begin{center}
\caption{Fixed-effect modeling on a heterogeneous set of patients and its average prediction residuals for each patient. }
\label{fig:fixedeffectproblem}
\includegraphics[scale=0.3]{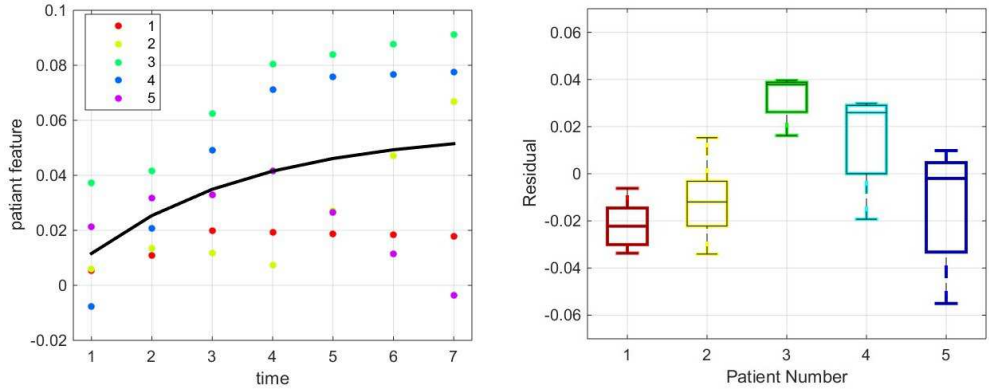}
\end{center}
\end{figure}

To better understand the above challenge, here we first apply a fixed-effect model (i.e. regression) on a data including patient with slightly different trends and shapes. Assume that the dataset consists of five patients each of which having 6 time-stamps corresponding to 6 extracted feature. Fig. \ref{fig:fixedeffectproblem} shows this data with its best polynomial fitted curve in the form of $\theta_1T^2+\theta_2T+\theta_3$.

As can be seen, the fitted fixed-effect model is extremely inaccurate in estimation of individual patients' behavior. This low performance which is reflected in relatively high residual values at the right of Fig. \ref{fig:fixedeffectproblem}, is due to the patient-specific trend of the data. 

\begin{figure}
\begin{center}
\caption{Mixed-effect modeling and its low prediction residual to address the patient-specific behaviors. }
\label{fig:mixedeffectnutshell}
\includegraphics[scale=0.3]{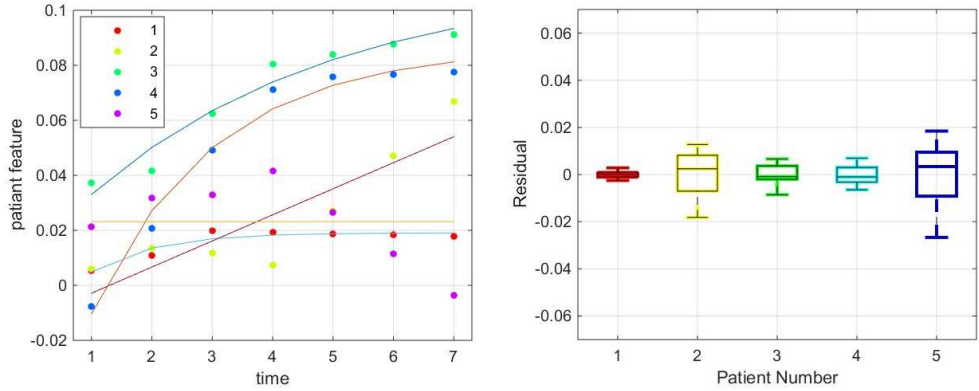}
\end{center}
\end{figure}

To address the above problem, mixed-effect modeling is used in this research. In a nutshell, the goal is to obtain one model per patient, as shown in Fig. \ref{fig:mixedeffectnutshell}. As shown, the flexibility of mixed-effect modeling allows us to estimate samples of each patient more accurately. 

Mixed-effect (ME) models are generalizations of linear or non-linear regression models for data that is collected and summarized in groups. ME models offer a flexible framework for analyzing grouped data while accounting for the within group correlation often present in such data.

\subsection{Parameter optimization}
After determining the model formulation, the next step is to estimate its parameters with respect to the input data. For this purpose, the Generalized Maximum Likelihood Estimation (GMLE) is used. In this method,  

\subsection{Predictor design}

The final step of the propose modelization algorithm is to generalize the trained model to estimate future state of a new sample (i.e. patient) with the similar diagnosis. In other words, given the medical images of a new patient, we aim at using the pattern in the available samples from past patients and fit it on the extracted features from new patient. The specification of such pattern can simply be stored in the model parameters that we already optimized in the previous step.

\begin{figure}
\begin{center}
\caption{Diagram of the proposed mixed-effect model for feature prediction, including the training phase (top part) and prediction phase (bottom part).}
\label{fig:predictordiagram}

\includegraphics[scale=0.1]{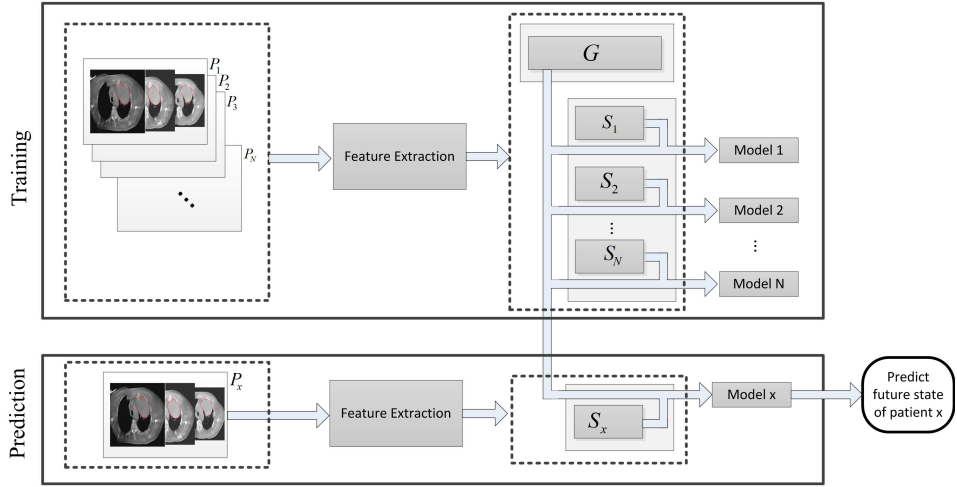}
\end{center}
\end{figure}

Fig. \ref{fig:predictordiagram} visualizes the predictor design and its relation with the trained mode. As can be seen, the diagram is split in two phases of training and prediction.

The training phase includes the offline process of parameter optimization. The outcome of this phase is one fixed-effect (i.e. $G$ or general model) and several random-effects (i.e. $S_i$ or specified models) for all patients in the dataset. The combination of each random-effect and the shared fixed-effect would provide the mixed-effect model of the patient the the used random-effect belongs to. 

With the same principle, any new patient can also be modeled in the prediction phase. Once a new patient $P_x$ arrives, its features are extracted exactly the same as the training phase. Then the extracted features are added to the existing feature set of the past patient to update the fixed-model $G$. This would also result in producing the random-effect of the new patient $S_x$. Combining the updated $G$ and $S_x$ provides a new mixed-effect model enabling the patient $P_x$ to predict its future state.

% =======================================
% =======================================
% =======================================
% =======================================
% =======================================
% =======================================

\section{Results}
\label{sec:results}
In this section, the performance of the proposed mixed-effect modeling of the features is presented. For this purpose, all the introduced features are extracted and trained by the proposed model. The primary results of the experiments showed that only the below three features produce relevant models: 
\begin{enumerate}
\item Volume
\item Mean of Jacobian matrix
\item Variance of Jacobian matrix
\end{enumerate}
\subsection{One-Leave-Out cross-validation}
Cross-validation is a model validation technique for statistical generalizability of a trained model on an independent data set. In the context of this research, we intend to measure the prediction accuracy of different models, trained by both an existing patients and the available data from a new patient. In other words, the goal is to measure how well the trained model will generalize and predict future state of a new patient, given his/her history that supposedly follows the same pattern as the the existing dataset.

The main objective of all cross-validation schemes is to guarantee that the evaluation step is performed with absolutely no bias and in a fair condition. These methods share an itertive step that partitions the dataset into two subsets:
\begin{itemize}
\item Training set: this partition is used to train a new system with the same training algorithm that we aim to validate.
\item Test set: this partition that usually consists of fewer number of samples from the dataset than the training set, is used to evaluate the performance of the system trained on the training set. 
\end{itemize}

\begin{figure}
\begin{center}
\caption{Leave-one-out cross-validation scheme for evaluating prediction performance of the trained model. In each iteration, one sample (the gray sample) is taken out of consideration and the the training algorithm is performed with the remaining samples. Then the prediction accuracy is evaluated on the removed sample.}
\label{fig:leaveoneout}
\includegraphics[scale=0.35]{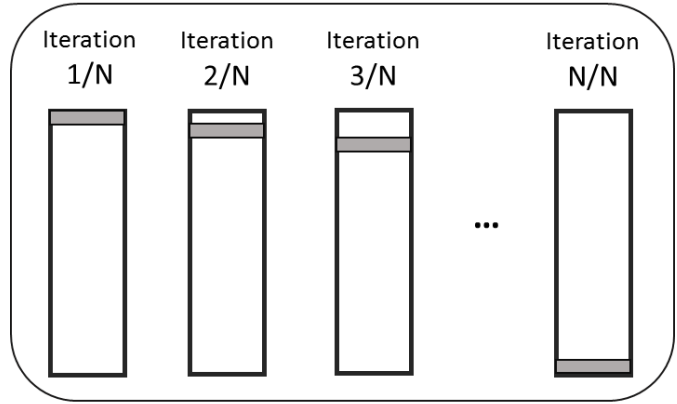}
\end{center}
\end{figure}

In the current leave-one-out method, the above partitioning step in each iteration puts one sample in the test set and all other samples in the training set. This is mostly due to the strict limitation on the number of patients available in the dataset. Fig. \ref{fig:leaveoneout} visualizes this scheme.

\subsection{Linear fixed effect vs. linear mixed effect}
\begin{figure}
\begin{center}
\caption{The shape of linear mixed-effect and liner fixed-effect modeling for the ``volume'' feature compared to the actual values of features.}
\label{fig:subplotLinear}
\includegraphics[scale=0.25]{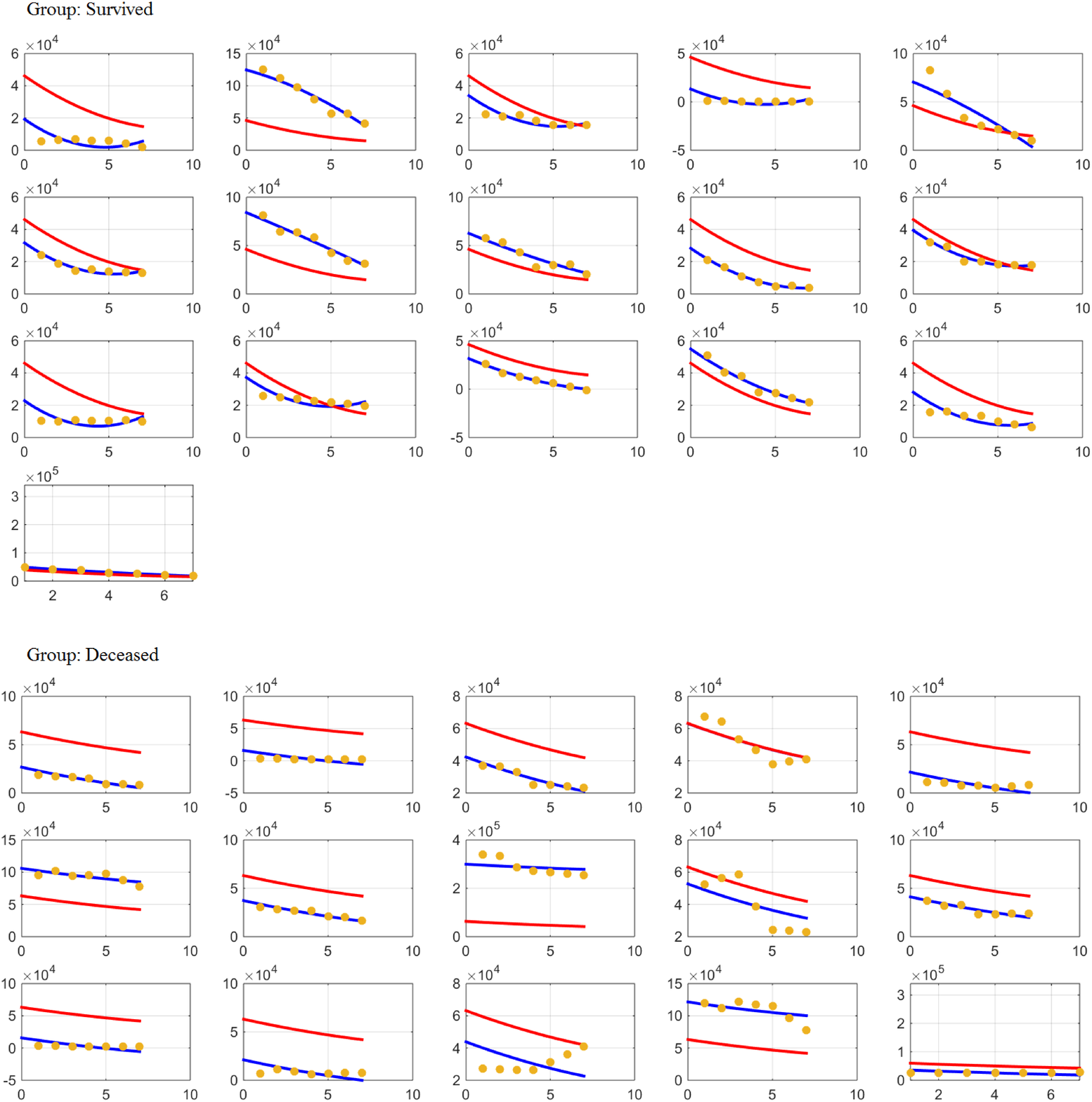}
\end{center}
\end{figure}

Fig. \ref{fig:subplotLinear} shows the modelization performance of fixed-effects compared to mixed-effect. In this figure, for both fixed-effect and mixed-effect modelization we have used linear equation in the form of $\theta_1T+\theta_2$ and the optimization step consisted of estimating $\Theta=[\theta_1,\theta_2]$. Compared to the actual feature values in different time-stamps $T$, it is clear that the linear mixed-effect modelization outperforms linear fixed-effect modelizaion for the given data. 

\subsection{Non-linear fixed effect vs. non-linear mixed effect}
\begin{figure}
\begin{center}
\caption{The shape of non-linear (polynomial) mixed-effect and non-linear fixed-effect modeling for the ``volume'' feature compared to the actual values of features.}
\label{fig:subplotNonLinear}
\includegraphics[scale=0.25]{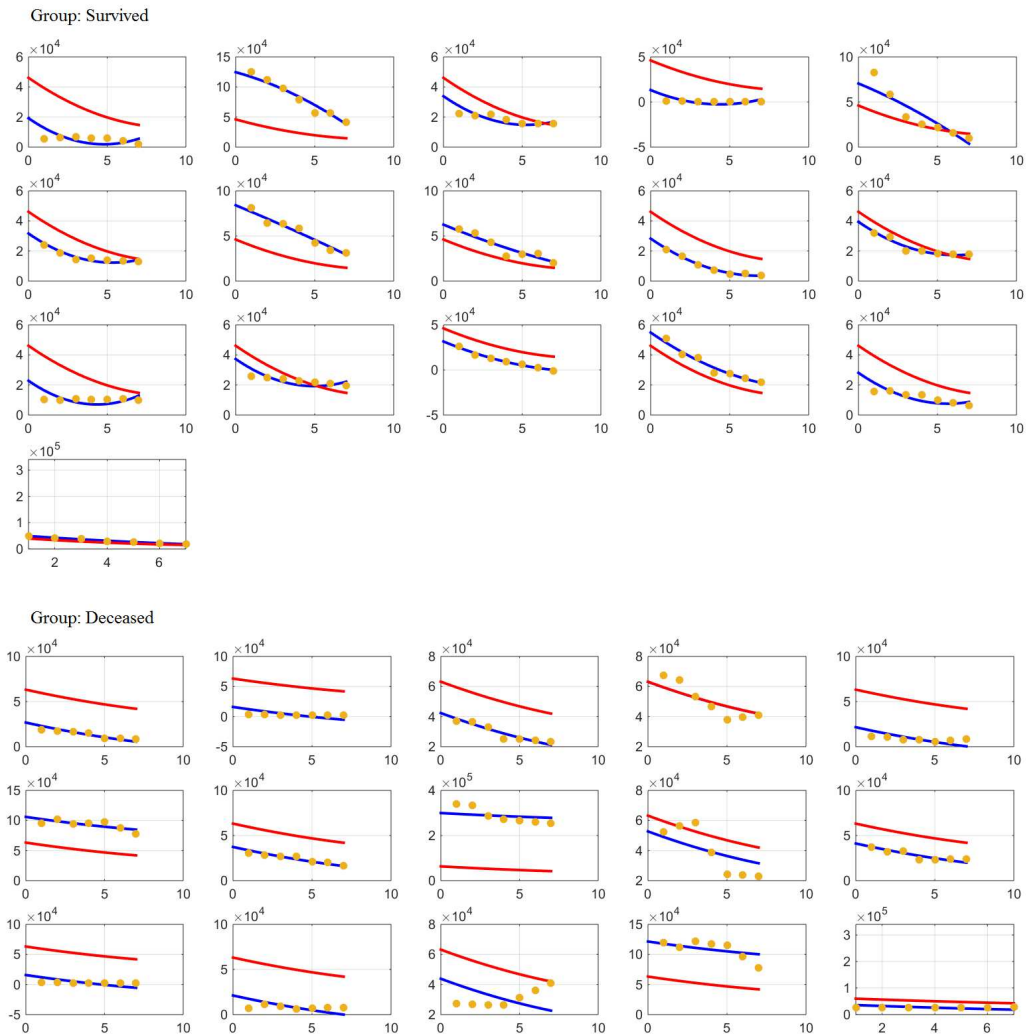}
\end{center}
\end{figure}

Fig. \ref{fig:subplotNonLinear} demonstrates a similar comparison as the previous one, with one difference. In this figure, non-linear polynomial formulation of $\theta_1T^2+\theta_2T+\theta_3$ was used. Correspondingly, the optimization step consisted of estimating the optimal parameter set $\Theta=[\theta_1,\theta_2,\theta_3]$. Similar to linear modeling, in the non-linear modelization too, the mixed-effect modelization has superior performance than fixed-effect modelization.

\subsection{Prediction accuracy}
Neither of Fig. \ref{fig:subplotLinear} and Fig. \ref{fig:subplotNonLinear} quantitatively measure the prediction performance of the proposed modelization. For this purpose, the cross-validation scheme was applied on the dataset. 

Three predictors are considered for the performance measurements:
\begin{enumerate}
\item Mixed-effect model with polynomial formulation: This predictor models the two groups of ``survived'' and ``deceased'' are separately. 
\item In-class fixed-effecd with polynomial formulation: This predictor considers two non-linear fixed-effect models corresponding to two groups and predicts each samples with the model of its corresponding group.
\item Out-class fixed-effect with polynomial formulation: The goal of using this predictor is to 
\end{enumerate}
\begin{figure}
\begin{center}
\caption{Prediction accuracy of mixed-effect modeling vs. fixed-effect modeling for the volume feature.}
\label{fig:predaccuracyvolume}
\includegraphics[scale=0.3]{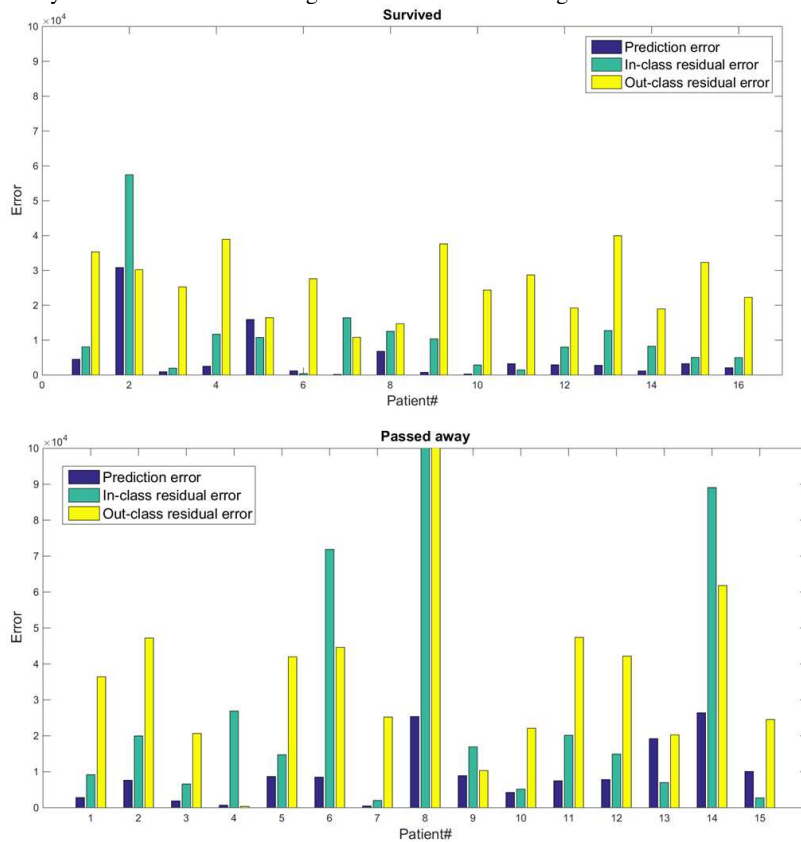}
\end{center}
\end{figure}

\subsection{Prediction accuracy feature 2: Mean of the Jacobian matrix}
\begin{figure}
\begin{center}
\caption{Prediction accuracy of mixed-effect modeling vs. fixed-effect modeling for the mean feature.}
\label{fig:predaccuracymean}
\includegraphics[scale=0.4]{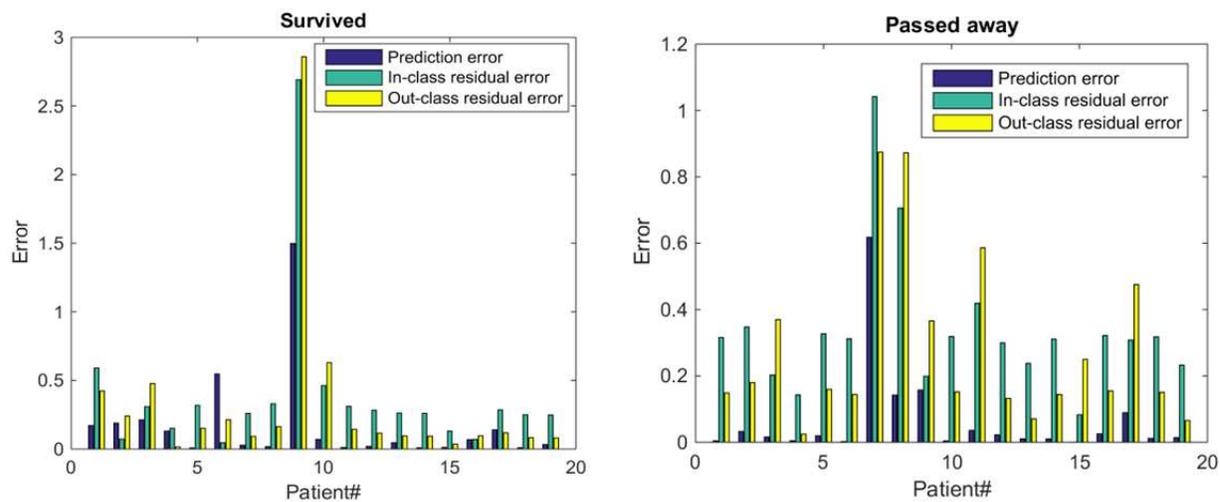}
\end{center}
\end{figure}

\subsection{prediction accuracy feature 3: Variance of the Jacobian matrix}
\begin{figure}
\begin{center}
\caption{Prediction accuracy of mixed-effect modeling vs. fixed-effect modeling for the variance feature.}
\label{fig:predaccuracyvar}
\includegraphics[scale=0.4]{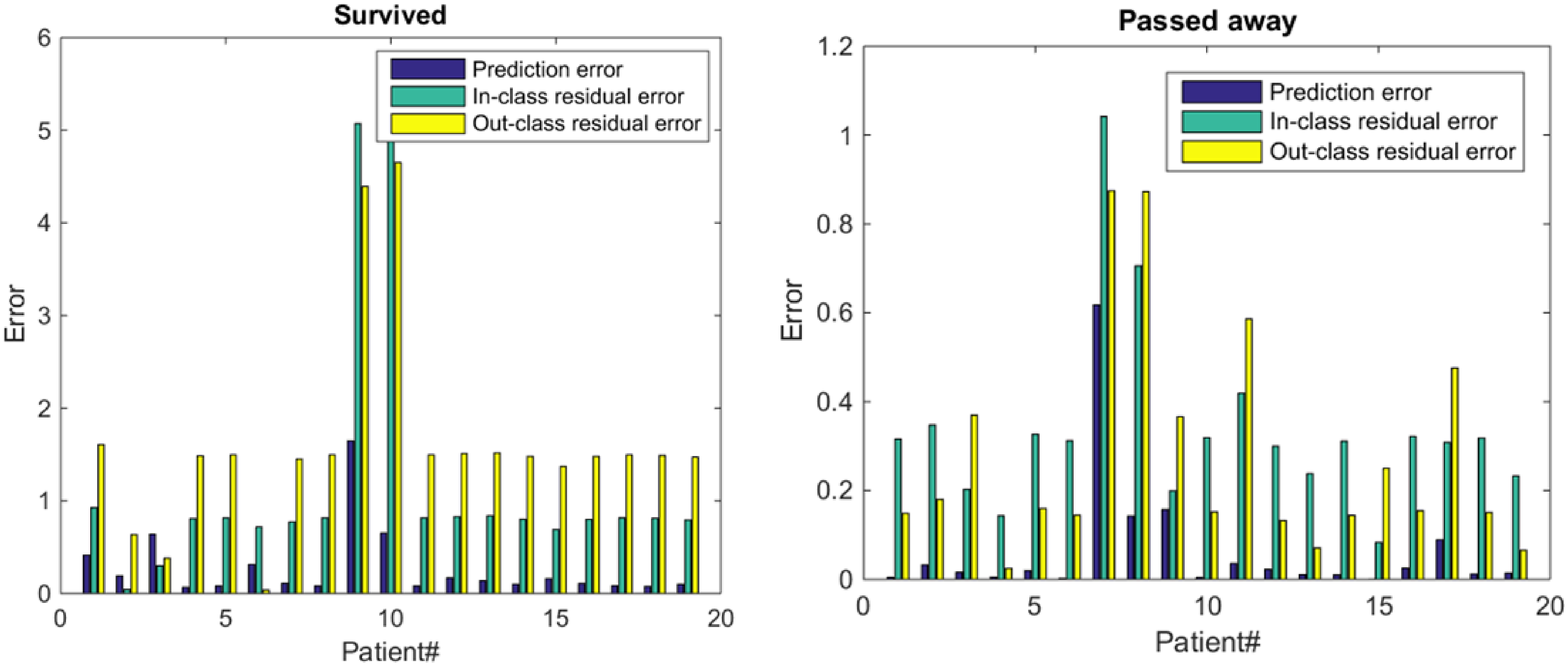}
\end{center}
\end{figure}

% =======================================
% =======================================
% =======================================
% =======================================
% =======================================
% =======================================

\section{Conclusion}
\label{sec:conclusion}
In this paper, different aspects of longitudinal modelization was reviewed. The ultimate goal was to optimize a model on a given dataset from a specific longitudinal feature extracted from past patients in a way that is capable of generalization for new patients with the same disease. The specific disease of this study was the lung cancer and the longitudinal features were volume of tumor and a set of features based on the deformation of tumor through the time.

The dataset that was used included two groups of patient, each having 19 members. All patients of these two groups have suffered from lung cancer and have undergone the same treatment that included 8 weeks of regular medical imaging. The difference between the two groups was that the patients of the first group survived the treatment and significantly reduced their tumor size. However, the patients of the second group did not succeed and passed away during or at the end of this period, despite tumor size reduction.

For the feature modelization, we proposed a mixed-effect modelization scheme that allows the modelization of each patient to benefit both from its patient-specific model and the general patient-independent model. The experiments showed that such modelization results in much higher prediction accuracy compared to fixed-effect modelization.

\bibliographystyle{unsrt}
\bibliography{myBib} 
\end{document}